# IT services design to support coordination practices in the Luxembourguish AEC sector.

Sylvain Kubicki[1], Annie Guerriero[1,2], Damien Hanser[1] and Gilles Halin[2]

[1] Centre de Recherche Public Henri Tudor. 29, avenue JF Kennedy
L 1855 - Luxembourg-Kirchberg
[2] MAP-CRAI - Research Centre in Architecture and Engineering.
2, rue Bastien Lepage. 54001 Nancy, France
{sylvain.kubicki, annie.guerriero, damien.hanser}@tudor.lu - gilles.halin@crai.archi.fr

**Abstract.** In the Architecture Engineering and Construction sector (AEC) cooperation between actors is essential for project success. The configuration of actors' organization takes different forms like the associated coordination mechanisms. Our approach consists in analyzing these coordination mechanisms through the identification of the "base practices" realized by the actors of a construction project to cooperate. We also try with practitioners to highlight the "best practices" of cooperation. Then we suggest here two prototypes of IT services aiming to demonstrate the value added of IT to support cooperation. These prototype tools allow us to sensitize the actors through terrain experiments and then to bring inch by inch the Luxembourgish AEC sector towards electronic cooperation.

## 1 Introduction

Cooperation between actors is essential for the success of a construction project. The short-lived groups of actors, heterogeneity of stakeholders and intern strategies of their firms are the main specificities of the AEC[1] sector. In opposition to other industries the rationalization of work processes and their computerization are still low developed in the construction of buildings sector.

However this is not due to a delay or an archaism of the sector compared to "leading edge industries". Indeed, the diversity of projects and architectural realizations is added to the complexity of groups of actors and relations between them. In this context, the change of work method takes time, and stakeholders able to impose it don't exist. The Luxembourguish construction sector is not an exception and presents the same particularities as those of its European neighbours.

Then, the cooperative processes could be improved. In fact, delays and building defaults regularly appear on building construction sites. They are notably due to dysfunctions in cooperative processes that actors perform. These processes have to be improved in order to limit these risks. IT innovation is a way to support these changes in professional practices.

---
[1] Architecture Engineering and Construction

In Luxembourg the Information Technology Resources Centre for Building (CRTI-B[2]) is an inter-professional organization, created in 1990. At the national level, the CRTI-B aggregates the main actors of the building sector: owners, architects, engineers, contractors etc. This organization supports integration of new Information and Communication Technologies in the building sector through innovation R&D projects. The overall objective of these projects is to lead tasks as closely as possible to the sector in order to propose concrete solutions (methods and tools software) to coordination needs of professionals coming from this working field (architects offices, design offices, home-building companies…). The primary goal of the Build-IT project is to enhance the competitiveness and the quality of the production process in the building sector by the use of ICT. Within the framework of this project, we focus on the practices of the exchange and the share of information that will ensure the interoperability between the actors of the Luxembourguish building sector. The Build-IT project encompasses a variety of research and development initiatives, most of which involve practitioners.

This article describes the first results of this project. First on a theoretical plan we address an analysis of actors' organizations in order to characterize coordination practices in building project. We present then two developments of IT services responding directly to the problems observed with practitioners, and the first validation elements. Finally, we conclude through opening future ways of actions to develop in the next stages of the project.

## 2    Cooperation processes in AEC projects

The terrain action carried out in the framework of the Build-IT project is completed by a theoretical background. Academic PhD works[3] reinforce this approach by characterizing and modelling cooperation and coordination processes in AEC.

### 2.1    Organization of actors and coordination mechanisms

In AEC projects, cooperation is extremely important because projects bring together numerous independent actors during short periods. Their activities are low predictable and they very often have to adapt their tasks and decisions to the specific problems they have encountered. Organization of actors takes different forms in this evolving context [1]. It is "hierarchical" when an actor is responsible of the work of the others [2,3] (i.e. building construction coordinator). We call it "adhocratic"[4] when actors are grouped in an informal way to solve a specific problem, punctual and unanticipated.

---

[2] http://www.crtib.lu
[3] Three PhD theses are and have been achieved in the Architecture and Engineering Research Centre (CRAI) at the Architecture School of Nancy, France (http://www.crai.archi.fr)

These two fundamental forms of actors' organizations coexist during the design and building construction phases.

Coordination of activities depends on these organization forms. In the "hierarchical organization" a coordinator monitors tasks progression, anticipates problems and organizes their solving. His work is based on specific documents and tools [5] helping him to diffuse coordination information, such as construction planning or meeting report. In "adhocratic organization", coordination is essentially informal, based on awareness of the others and situated action [6]. It is an essential coordination mode, e.g. during the building construction activity. It ensures adaptability of the actions to the unpredictability of the activity and to frequent changes. In this coordination form, documents given by hierarchy don't serve directly the actions of the actors. They provide contextual information that actors need to adapt their decisions.

### 2.2 Cooperative processes and practices

We have described the actors' organizations and coordination mechanisms associated existing in construction projects. Then coordination tasks are essential activities. Indeed, the AEC sector involves heterogeneous teams and activities not really predictable. Cooperative processes realized in a construction project are not precisely defined. However, a certain number of practices exist and assist the cooperation between actors.

So, our approach doesn't consist in defining unique processes, repeatable or standardized. To the contrary, we try, with professionals, to highlight daily practices that can encourage and improve the cooperation in the construction projects.

In this approach, we inspire about methods of processes assessment and continuous improvement in the organizations, such as ISO 15504 (SPICE[4]). An ISO 15504 assessment consists in selecting a certain number of business processes (Process Assessment Model) in order to evaluate their maturity with people implicated in their realization. Each process is analyzed according to the "base practices" that allow its accomplishment. We suggest to apply this processes/practices division in order to tackle the cooperation from a "business" viewpoint. That is why we identify practices during interviews with professionals of the construction sector from Luxembourg.

We will see now that cooperation practices are directly linked with organizational configurations and also, coordination mechanisms.

### 2.3 IT services supporting these practices

Many IT tools exist and assist actors during the execution of these cooperative practices. To manage coordination, the coordinator uses planning tools. The building construction meeting report informs about the state of the construction activity at a given moment. It is written after each meeting and regroups in a document, which will be validated by all the participants, all the decision taken, identified problems (more and more often illustrated with some pictures), state of the progress and other pieces of information [7]. More recently 4D CAD tools consist in an interface that shows relation between the 3D mock-up and the execution planning [8, 9]. The objective of such tools is to simulate the state of the construction activity. Moreover, it improves considerably communication with the owner and it allows to ripen the execution planning.

---

[4] SPICE project official website: http://www.sqi.gu.edu.au/spice/

These tools, which we have identified above, have a real utility in the construction activity coordination. They inform about the construction process, about the state of the activity and its execution. However, their use is not really common. A certain number of blocking factors explain it (e.g. tools appropriation, changes relative to the method of work, organizational changes, etc.). In the Luxembourguish construction sector, this problem exists like elsewhere. The privileged place of the CRTI-B allows us to regroup numerous actors representing the different professions of the construction sector around the Build-IT project, and to think together about real needs of the digital cooperation.

### 2.4 Problem and hypotheses

In the framework of the Build-IT project, we focus on the identification of the essential cooperative processes and on their explaining with the professionals themselves during "Working Groups". This "applied approach" finds its origin in different works relative to the organizations and the coordination mechanisms. The identification, with the actors, of coordination practices (essential or problematic) allows us to suggest IT services supporting business needs formulated by the sector practitioners. Table 1 suggests a synthetic and non-exhaustive view of this approach. It puts in relation organizations and coordination mechanisms with the coordination practices identified as essential and IT services that we suggest to support them.

**Table 1.** Organization, coordination practices and IT services

| Configuration of the organization | Coordination mechanism | Coordination practices | Associated IT service |
|---|---|---|---|
| Hierarchical configuration | Direct supervision | Meeting report writing | Meeting report management service |
| | | Meeting report consultation | |
| | | Reaction on a remark | |
| | | Plans structuring | Plans management service |
| | | Plans update | |
| | | Plans annotation | |
| | | Notification of published plan | |
| | | Diffusion monitoring | |
| | | Exchange traceability | |
| Adhocratic configuration | Mutual adjustment | Awareness practices | Context perception support service |
| | | Consultation of various documents | |

## 3   IT services development to support coordination practices

The underlining of the cooperation practices, and the development of a model of the cooperation context [10] lead us to envisage their support in the form of a coherent set of IT services adapted to the needs of practitioners.

These needs often relates to the projects, their sizes, characteristics of the teams, types of the contract, etc. Our approach consists in considering modular services (one independent from each other) and in managing the exchange of information between service-specific HCI (Human-Computer Interfaces).

### 3.1   Meeting report writing and consultation service

The first Build-IT service[5], developed in the Build-IT project, is intended to manage exchanges around the meeting report. Then it supports direct supervision in hierarchical organizations. It is a typical situation of construction activity, where the coordinator writes a meeting report describing particular points to be adjusted.

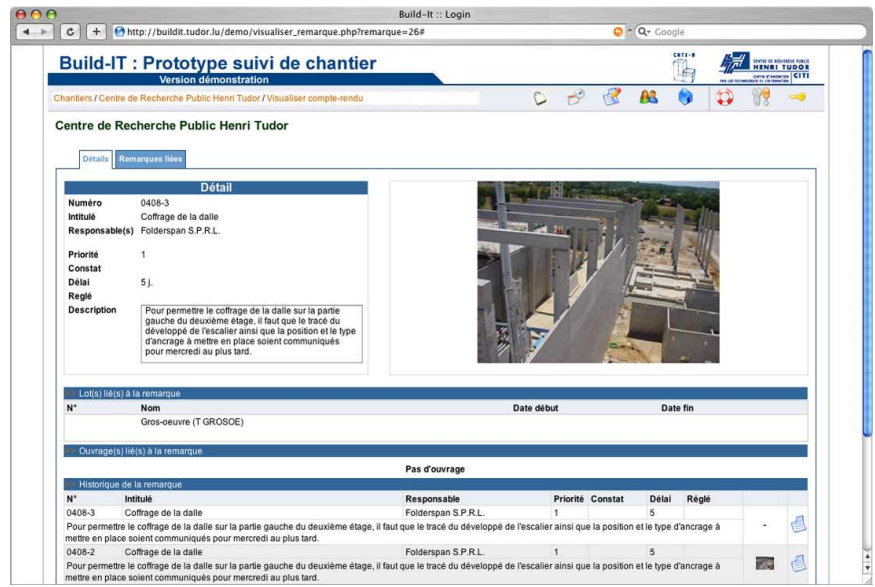

**Fig. 1.** Built-IT prototype screenshot

Our analysis of building activity meeting report and processes linked to this document allowed us to identify firstly, components of the meeting report (e.g. Presence and diffusion list, progress, list of remarks, etc.) and secondly, to determine services associated to the business practices.

---

[5] http://buildit.tudor.lu/demo - Login: "demo" - Password: "demo"

The meeting report prototype integrates three services. *Writing service* covers functionalities intended to the writer. The prototype guides the writing by using forms. These forms correspond to generic components of meeting reports identified during the analysis phase. *Dynamic consultation service* covers functionalities of search. The tool offers the possibility to combine information filters to find easily information the user needs (filter on responsible people, lots, building elements…). Moreover, the search in three levels allows to restrict gradually the field of search: a first search level within various current construction sites, a second search level inside a construction site and finally, the last one, inside the meeting report. Finally r*eaction service* covers functionalities intended to react to a remark. The tool allows the reader to react to a remark if he feels that its content is erroneous or requires further information. The centralization of information and the traceability of exchanges linked to the meeting report inside the tool is a way, on one hand, to enhance coordination between various contributors and on the other hand, to identify more easily the source of problems.

Currently, the experiment of this tool is in progress. It is used in 8 real construction projects. The experiment will allow us to verify the relevance of the tool in real situations of building activity, the consistency of visualized data, the usability of end-users interfaces and the appropriation of the tool by practitioners.

### 3.2 Bat'iViews: a context perception support service

Information related to coordination is represented in numerous views attached to documents, coordination tools or communication tools. Present practices consist in finding related pieces of information in the diverse useful documents, i.e. meeting report, planning and others. In terms of coordination, the need is to support practices of mutual adjustment. These practices are observed in unanticipated situations, in which the actors have to auto-coordinate. Concretely, the quality of this coordination depends on the capacity to obtain a global vision of the problem to be resolved, and to envisage risks that some potential solutions present. To improve context comprehension by the actors we think it is necessary to provide a representation, showing relations existing between the different elements of the context.

Bat'iViews prototype [11] suggests to make use of views manipulated everyday by the construction stakeholders and to integrate them in a navigation tool showing relations existing between content elements of each one. We choose 4 dynamic coordination views to develop the prototype: meeting report view, planning view, 3D mock-up view and a view of all remarks in all meeting reports. In order to show relations between elements of different views, the tool is based on the multi-visualization principle [12, 13]. It provides different views' arrangements to the user allowing him to navigate in the project context. The concepts to link through the views depend on the model of each view: i.e. meeting report displays "remarks" concerning "actors" and "building element", planning shows "tasks" and 3D mock-up represents "building objects". User-interaction is generated by the selection of one of these elements in each view. It consists in finding the corresponding concepts in the other views models and to highlight them. We call it a "free navigation": each view can generate interaction.

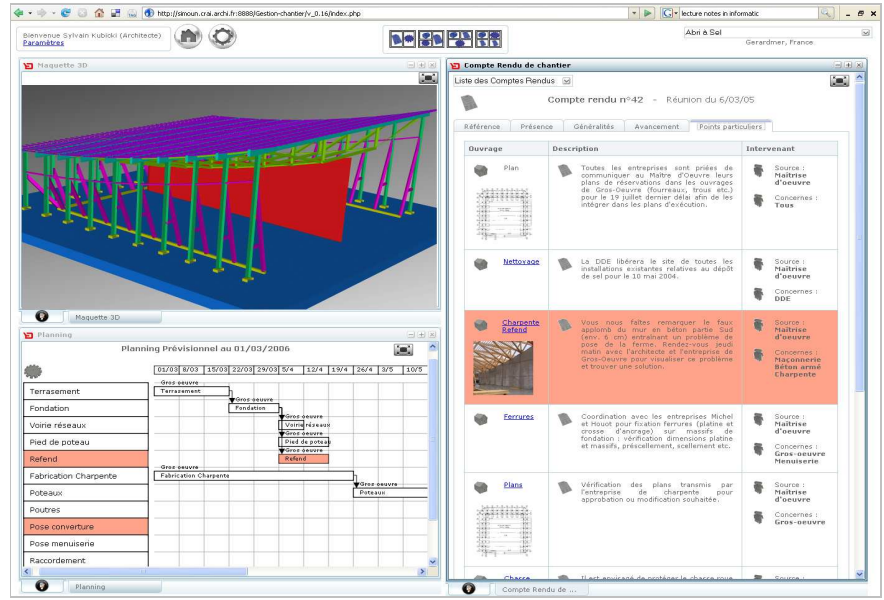

**Fig. 2.** Bat'iViews prototype screenshot

Figure 2 illustrates an arrangement in Bat'iViews[6]. It is composed of three views: 3D mock-up, planning and meeting report. An element selected in the 3D mock-up (here the main wall in red) is linked with related concepts in the planning (construction task of this wall) and a remark in the meeting report (e.g. There is a problem of synchronization between wall construction and roof frame construction).

## 4 Conclusion and future works

The works presented here take place in the Build-IT project, which aims at guiding the Luxembourguish construction sector towards digital cooperation. The hypothesis that we argue is that numerous IT tools exist but their use is weak. The reasons are multiple: these tools are not really adapted to the needs of a particular industrial sector, and even more the actors don't see a real value-added in their use. In this context, the objective of the project is to lead actions of sensitization and service developments. This article describes the first steps of this action. We highlight potentialities of services managing coordination information about the building construction activity (meeting report service) and also services improving information understanding (contextual multi-visualization service). The experiments realized with those tools reinforce the hypothesis that if they are designed in collaboration with professionals, their appropriation and transfer to the sector is easier.

---

[6] http://www.crai.archi.fr/bativiews

The next step is relative to the processes of plans exchange. It is about processes implicating the totality of stakeholders of the project, because sending and reception of documents concern everybody. So, we consider them like being essential in the hierarchy as in the adhocracy, and their coordination recovers from direct supervision and also from mutual adjustment. We have currently identified some base practices through a set of interviews with professionals (Cf. Table 1). We are now generalizing them to all the actors. Working Groups allow us to discuss and exchange with the professionals in order to highlight a set of "best practices". They will lead us then to suggest a set of IT services in the form of a prototype implementing these best practices. This demonstrator will allow itself to generalize the sensitization actions for the sector. The future steps of the Build-IT project are just drafted in the form of coordination scenarios. Beyond the management of documents of the project, the sector is going towards the "common and shared object" artefact. That will proceed certainly at first by the introduction of a common reference ontology, enabling to describe building elements through daily documents. This step will be followed by the generalization of the Building Information Model describing geometrically building elements and allowing everyone to add information in function of his particular point of view.